\documentclass[nofootinbib, reprint, preprintnumbers, amsmath, amssymb, amsfonts, aps, pra, superscriptaddress, floatfix]{revtex4-2}
\usepackage{mathtools}
\usepackage{graphicx}
\usepackage{bm}
\usepackage{color}
\usepackage[normalem]{ulem}
\usepackage{silence}
\usepackage[colorlinks=true, linkcolor=blue, citecolor=blue, urlcolor=blue]{hyperref}

\begin{document}
\title{Solving Differential Equations Using Continuous-Variable Quantum Annealing}

\author{Kazuki Miyanishi}
% \email{a21.j7j3@g.chuo-u.ac.jp}
\affiliation{Department of Electrical, Electronic, and Communication Engineering, Chuo University, 1-13-27 Kasuga, Bunkyo-ku, Tokyo, 112-8551, Japan}

\author{Soshun Naito}
\email{naito@biom.t.u-tokyo.ac.jp}
\affiliation{Department of Information and Communication Engineering, Graduate School of Information Science and Technology, The University of Tokyo, 7 Chome-3-1 Hongo, Bunkyo, Tokyo, 113-8654, Japan}

\author{Asuka Koura}
\affiliation{Department of Electrical, Electronic, and Communication Engineering, Chuo University, 1-13-27 Kasuga, Bunkyo-ku, Tokyo, 112-8551, Japan}

\author{Kazue Kudo}
\affiliation{Department of Computer Science, Ochanomizu University, 2-1-1 Ohtsuka, Bunkyo-ku, Tokyo 112-8610, Japan}

\author{Toshiki Yamaji}
\affiliation{National Institute of Advanced Industrial Science and Technology, 1-1-1 Umezono, Tsukuba, Ibaraki 305-8568, Japan}

\author{Yuichiro Matsuzaki}
\email{ymatsuzaki872@g.chuo-u.ac.jp}
\affiliation{Department of Electrical, Electronic, and Communication Engineering, Chuo University, 1-13-27 Kasuga, Bunkyo-ku, Tokyo, 112-8551, Japan}

\date{\today}

\begin{abstract}

Most existing quantum annealing approaches are formulated for qubit-based architectures. Consequently, applying them to continuous-variable optimization problems requires discretizing the variables, which can incur substantial qubit overhead. Continuous-variable quantum annealing based on bosonic systems has recently been proposed as an alternative framework, in which each optimization variable is directly encoded in a bosonic mode, such as a cavity mode. In this work, we develop a continuous-variable quantum annealing formulation for solving linear differential equations. By recasting the determination of the solution as a continuous-variable optimization problem, the differential equation can be mapped onto an objective function compatible with bosonic quantum annealing. Numerical simulations of second-order linear differential equations demonstrate that, under the conditions considered, the proposed formulation reproduces the corresponding analytical solutions. These results establish a potential route toward solving differential equations without the discretization overhead inherent in qubit-based implementations.

\end{abstract}

\maketitle

\section{Introduction}
Quantum annealing (QA) is a method for finding the ground state of an Ising Hamiltonian \cite{Kadowaki_1998,finnila1994quantum,farhi2000quantum,farhi2001quantum,apolloni1989quantum}.
Because many combinatorial optimization problems can be written in the form of an Ising Hamiltonian, QA has attracted attention in recent years as a new approach to solving combinatorial optimization problems \cite{Lucas2014,Hauke2020}.
QA is performed by slowly changing the Hamiltonian of a system.
The initial Hamiltonian is called the driver Hamiltonian and has a trivial ground state.
The final Hamiltonian is called the problem Hamiltonian and is the Ising Hamiltonian that represents the optimization problem.
By preparing the initial state of the system as the ground state of the driver Hamiltonian and changing the Hamiltonian so as to satisfy the quantum adiabatic condition \cite{1390001204189531392,Jansen_2007,ambainis2006elementaryproofquantumadiabatic}, the final state can approach the ground state of the problem Hamiltonian.

Most previous studies on quantum annealing for combinatorial optimization have focused on qubit-based implementations \cite{dema2020support,date2021adiabatic}. However, because qubits are intrinsically two-level systems, they cannot directly represent continuous optimization variables. Continuous variables must therefore be discretized and encoded into registers of multiple qubits. A widely used encoding scheme is binary expansion, in which each variable is represented by a finite-length binary string \cite{chancellor2019domain,karimi2019practical,tamura2021performance,krakoff2022controlled,PhysRevA.108.042403}. In this approach, achieving higher numerical precision requires increasing the number of bits, and hence the number of qubits assigned to each variable. Consequently, the total qubit requirement can grow with the desired precision.

Recently, continuous-variable quantum annealing based on bosonic systems has been proposed as an alternative to conventional qubit-based implementations \cite{koura2024linear}. In this framework, bosonic modes, such as cavity modes, are employed as computational degrees of freedom, and the variables are represented by creation and annihilation operators.
The driver Hamiltonian is designed so that its ground state is the bosonic vacuum state, which can be prepared as the initial state of the annealing protocol. Because bosonic modes possess continuous degrees of freedom, this approach can represent continuous variables directly, thereby avoiding the qubit overhead associated with discretization and binary encoding. Linear regression has been investigated as a representative application of this framework \cite{koura2024linear}. However, its applicability to other classes of continuous-variable problems remains largely unexplored.

In this work, we investigate the solution of linear differential equations as a new application of continuous-variable quantum annealing. After discretization, a linear differential equation can be converted into a system of linear equations, which can subsequently be reformulated as a generalized eigenvalue problem or a quadratic optimization problem \cite{bravo2023variational,sato2023variational}. Based on such formulations, several quantum algorithms employing variational quantum circuits have been proposed \cite{PhysRevA.104.022418,PhysRevA.104.052409,liu2022application,demirdjian2022variational}. Quantum-annealing-based approaches have also been investigated \cite{Kudo_2026,PhysRevA.99.052355,PhysRevA.104.032426,Criado_2023}. In conventional qubit-based implementations, however, the continuous values of the discretized solution must be encoded using finite qubit registers, so that the required number of qubits increases with the desired numerical precision. In the present work, we instead formulate the determination of the solution to a linear differential equation as a quadratic optimization problem over continuous variables. The resulting cost function is mapped onto a bosonic Hamiltonian whose ground state encodes the solution of the discretized differential equation. This construction provides a direct framework for applying continuous-variable quantum annealing to linear differential equations.

This paper is organized as follows.
First, we explain the representation of variables and cost functions in continuous-variable QA.
Next, we show how a linear differential equation can be formulated as a quadratic optimization problem through time discretization and finite-difference approximation.
Finally, we evaluate the proposed method through numerical simulations of linear differential equations with known analytical solutions.

\section{Quantum Annealing}
Quantum annealing is a method for finding the ground state of an Ising Hamiltonian \cite{Kadowaki_1998,finnila1994quantum,farhi2000quantum,farhi2001quantum,apolloni1989quantum}.
The Hamiltonian used in QA is written as
\begin{equation}
  \hat{H}(t) = \frac{t}{T}\hat{H}_P + \left(1-\frac{t}{T}\right)\hat{H}_D
\end{equation}

where $\hat{H}_{\mathrm{P}}$ and $\hat{H}_{\mathrm{D}}$ denote the problem and driver Hamiltonians, respectively. The problem Hamiltonian is constructed such that its ground state encodes the solution of the optimization problem, whereas the driver Hamiltonian is chosen to have a known and readily preparable ground state. In QA, the total Hamiltonian is continuously varied from $\hat{H}_{\mathrm{D}}$ to $\hat{H}_{\mathrm{P}}$, with the system initially prepared in the ground state of $\hat{H}_{\mathrm{D}}$. According to the quantum adiabatic theorem \cite{1390001204189531392,Jansen_2007,ambainis2006elementaryproofquantumadiabatic}, when the Hamiltonian changes sufficiently slowly and the standard adiabatic conditions are satisfied, the evolving state remains close to the instantaneous ground state. Consequently, for a sufficiently long annealing time $T$, the final state approaches the ground state of $\hat{H}_{\mathrm{P}}$ and thereby provides a solution to the optimization problem.

\subsection{Representation of Variables and Cost Functions}
For $\widetilde{N}$ continuous variables $\bm{\theta} = (\theta_1,\theta_2,\cdots, \theta_{\widetilde{N}})^T$, assume that the cost function $L(\bm{\theta})$ can be written in the following quadratic form:
\begin{align}
    \label{QUBOcost}
    L(\bm{\theta})
    &= \bm{\theta}^T Q \bm{\theta} + \bm{r}^T \bm{\theta} \nonumber \\
    &= \sum_{i=1}^{\widetilde{N}} \sum_{j=1}^{\widetilde{N}} Q_{ij} \theta_i \theta_j + \sum_{i=1}^{\widetilde{N}} r_i \theta_i
\end{align}
Here, $Q$ and $\bm{r}$ are a real-valued coefficient matrix and vector, respectively.

In qubit-based QA, each variable $\theta_i$ is discretely represented as
\begin{align}
    \label{QUBOtheta}
    \theta_i = s_i + (t_i - s_i) \sum_{k=1}^{K} 2^{-k} x_{ik}
\end{align}
Here, $s_i$ and $t_i$ are the lower and upper bounds of $\theta_i$, respectively, and $x_{ik}$ is a binary variable.
The possible values of $\theta_i$ are distributed at equal intervals in the range $s_i \leq \theta_i < t_i$.
The parameter $K$ corresponds to the representation precision of the variable, and the required number of qubits increases in proportion to $K$.
Because this substitution expresses $\theta_i$ as a linear combination of the binary variables $x_{ik}$, applying it to all $\theta_i$ allows $L(\bm{\theta})$ to be represented as a QUBO (Quadratic Unconstrained Binary Optimization) problem \cite{chancellor2019domain,karimi2019practical,tamura2021performance,krakoff2022controlled}.

In continuous-variable QA using cavities, the variable $\theta_i$ and its products are represented by creation and annihilation operators as follows \cite{koura2024linear}:
\begin{align}
\label{QUBOconvert}
    \theta_i^2 &\to \hat{a}_i^\dagger \hat{a}_i\nonumber\\
    \theta_i \theta_j &\to \frac{1}{2} \left( \hat{a}_i^\dagger \otimes \hat{a}_j + \hat{a}_i \otimes \hat{a}_j^\dagger \right)\nonumber\\
    \theta_i &\to \frac{1}{2} \left( \hat{a}_i + \hat{a}_i^\dagger \right)
\end{align}
Here, $\hat{a}_i^\dagger$ and $\hat{a}_i$ are the creation and annihilation operators of the $i$-th cavity.
Applying this substitution to Eq.~\eqref{QUBOcost}, the cost function can be represented as the following problem Hamiltonian $\hat{H}_{\mathrm{P}}$:
\begin{align}
    \hat{H}_{\mathrm{P}}
    = &\sum_{i<j} \frac{Q_{ij} + Q_{ji}}{2} \left( \hat{a}_i^\dagger \otimes \hat{a}_j + \hat{a}_i \otimes \hat{a}_j^\dagger \right) + \sum_{i=1}^{\widetilde{N}} Q_{ii} \hat{a}_i^\dagger \hat{a}_i \nonumber \\
    &+ \sum_{i=1}^{\widetilde{N}} \frac{r_i}{2} \left( \hat{a}_i + \hat{a}_i^\dagger \right)
\end{align}

\section{Hamiltonian Representation of Linear Differential Equations}
In this paper, we represent a linear differential equation for a function $x(\tau)$ on the interval $\tau \in [T_0,T_1]$ in the following form:
\begin{equation}
    \sum_{n=0}^{N}A_n(\tau)\frac{d^n x}{d\tau^n} (\tau) = B(\tau)
\end{equation}
Here, $A_n(\tau)$ and $B(\tau)$ are functions that do not depend on $x(\tau)$.
Because continuous-variable QA cannot directly handle functions, time must be discretized.
In this paper, the interval $[T_0,T_1]$ is divided into $M$ equal subintervals.
The width of each interval is $\Delta\tau = (T_1-T_0)/M$, and the $i$-th time point is $\tau_i = T_0 + i\Delta\tau$ for $i=0,1,\cdots,M$.
Instead of directly determining the continuous function $x(\tau)$, we estimate the function values $x(\tau_i)$ at the discretized time points.
Then the differential equation at each time point is written as
\begin{equation}
    \label{eq:divid_linear_equation}
    \sum_{n=0}^{N}A_n(\tau_i) \frac{d^nx}{d\tau^n} (\tau_i) = B(\tau_i)\qquad(i=0,1,\cdots,M)
\end{equation}

The continuous-variable QA formulation used here is based on the optimization of quadratic functions.
Therefore, by approximating the derivative terms as linear combinations of $x(\tau_i)$, the square of the difference between the two sides of Eq.~\eqref{eq:divid_linear_equation} becomes a quadratic function of $x(\tau_i)$.
In this work, we use central finite differences as a concrete approximation.
Examples are
\begin{align}
    \label{sabun_origin}
    \frac{dx}{d\tau}     &\simeq \frac{x(\tau_{i+1})-x(\tau_{i-1})}{2\Delta \tau} \\
    \frac{d^2x}{d\tau^2} &\simeq \frac{x(\tau_{i+1})-2x(\tau_{i})+x(\tau_{i-1})}{\Delta \tau^2}\nonumber \\
    \frac{d^3x}{d\tau^3} &\simeq \frac{x(\tau_{i+2})-2x(\tau_{i+1})+2x(\tau_{i-1})-x(\tau_{i-2})}{2\Delta \tau^3}\nonumber \\
                         &\vdots\nonumber 
\end{align}

Applying the above approximations to the derivative terms in Eq.~\eqref{eq:divid_linear_equation}, we obtain
\begin{equation}
    \label{senkei_after_sabun}
    C\bm{x}-\bm{b}=\bm{0}
\end{equation}
Here,
\begin{align}
    \bm{x}
    =
    \left(
    x(\tau_0),
    x(\tau_1),
    \cdots,
    x(\tau_{M})
    \right)^T
\end{align}
is the vector at the grid points, $C$ is the coefficient matrix obtained by the finite-difference approximation, and $\bm{b}$ is the vector containing the constant terms.
The elements of $C$ depend on the coefficients $A_n(\tau)$ in Eq.~\eqref{eq:divid_linear_equation} and on the derivative order $n$ of the variable $x$.
The elements of $\bm{b}$ depend on the constant term $B(\tau)$ in Eq.~\eqref{eq:divid_linear_equation}:
\begin{align}
    \bm{b}
    =
    \left(
    B(\tau_0),
    B(\tau_1),
    \cdots,
    B(\tau_{M})
    \right)^T
\end{align}
The squared residual of Eq.~\eqref{senkei_after_sabun} is then
\begin{align}
L(\bm{x})
=
(C\bm{x}-\bm{b})^T
(C\bm{x}-\bm{b})
\label{eq:matrix_cost}
\end{align}
and expanding Eq.~\eqref{eq:matrix_cost} gives
\begin{align}
L(\bm{x})
=
\bm{x}^TC^TC\bm{x}
-
2\bm{b}^TC\bm{x}
+
\bm{b}^T\bm{b}
\label{eq:matrix_expand}
\end{align}
Defining
\begin{align}
Q=C^TC
\qquad
\bm{r}=-2C^T\bm{b},
\qquad
c=\bm{b}^T\bm{b}
\label{eq:Q_definition}
\end{align}
we obtain
\begin{align}
L(\bm{x})
=
\bm{x}^TQ\bm{x}
+
\bm{r}^T\bm{x}
+
c
\label{eq:quadratic_cost}
\end{align}
Thus, the problem can be formulated as a cost function in the same quadratic form as Eq.~\eqref{QUBOcost}.

The formulation described above can be applied to arbitrary finite-order linear ordinary differential equations.
To uniquely determine the solution of a differential equation, appropriate initial and terminal conditions must be imposed according to the order of the equation.
In the numerical experiments in this paper, as a basic example of this general framework, we consider second-order linear ordinary differential equations with initial and terminal conditions.

\section{Numerical Evaluation}

\subsection{Experimental Setup}
\label{sec:setting}
In the numerical experiments, we consider the following two linear differential equations:
\begin{align}
    \label{eq:example_1}
    \frac{d^2x}{d\tau^2} &= 2, \\
    \label{eq:example_2}
    \frac{d^2x}{d\tau^2}+5\frac{dx}{d\tau} &= -1
\end{align}
Equation~\eqref{eq:example_1} contains only a second derivative, whereas Eq.~\eqref{eq:example_2} contains both first and second derivatives.
In both cases, $x(0)=0$ and $x(1)=1$ are given as the initial and terminal conditions.
Unless otherwise stated, the number of divisions is set to $M=7$, the step size is $\Delta\tau=1/M$, and the six values $x(\tau_1),x(\tau_2),x(\tau_3),x(\tau_4),x(\tau_5),x(\tau_6)$ are estimated.
Because the two endpoints are fixed, the required number of cavities is $M-1$.

As the driver Hamiltonian $\hat{H}_{\mathrm{D}}$ for continuous-variable QA, we use the following Hamiltonian whose ground state is the vacuum state:
\begin{align}
    \label{Hd}
    \hat{H}_{\mathrm{D}} &= d\sum_{i=1}^{M-1} \hat{a}_i^\dagger \hat{a}_i^{}
\end{align}
Here, $d$ is a coefficient that adjusts the strength of $\hat{H}_{\mathrm{D}}$, and unless otherwise stated we set $d=1.0$.
The initial state is chosen as the vacuum state, which is the ground state of $\hat{H}_{\mathrm{D}}$.
Unless otherwise stated, the annealing time is set to $T=10000$.
After continuous-variable QA, the value of $x(\tau_i)$ is obtained by evaluating the expectation value of the operator $\hat{x}_i=\frac{1}{2}(\hat{a}_i+\hat{a}_i^\dagger)$ corresponding to each cavity.
These numerical simulations were implemented using QuTiP \cite{Qutip,qutip5}.

\subsection{Application to \texorpdfstring{$\displaystyle \frac{d^2x}{d\tau^2}=2$}{d2x/dtau2 = 2}}
Using Eq.~\eqref{sabun_origin} to approximate the second derivative in Eq.~\eqref{eq:example_1} as a linear combination of $x(\tau_i)$, we obtain, for each internal point $i=1,\cdots,M-1$,
\begin{align}
    \frac{x(\tau_{i+1})-2x(\tau_i)+x(\tau_{i-1})}{(\Delta \tau)^2} = 2 
    \label{after_sabun_1}
\end{align}
This system of linear equations is converted into the cost function described in Section~3.
Then the matrix and vector in Eq.~\eqref{senkei_after_sabun} are given by
\begin{align}
    \bm{x}
    =
    \left(
    x(\tau_1),
    x(\tau_2),
    \cdots,
    x(\tau_{M-1})
    \right)^T
\end{align}
\begin{align}
C_{ij}
=
\begin{cases}
-\dfrac{2}{(\Delta\tau)^2},
&
i=j,
\\[2mm]
\dfrac{1}{(\Delta\tau)^2},
&
|i-j|=1,
\\[2mm]
0,
&
\mathrm{otherwise},
\end{cases}
\label{eq:A_matrix}
\end{align}
and
\begin{align}
\bm{b}
=
\left(
2-\frac{x(\tau_0)}{(\Delta\tau)^2},
2,
\cdots,
2,
2-\frac{x(\tau_M)}{(\Delta\tau)^2}
\right)^T
\label{eq:b_vector}
\end{align}
By applying the transformation from Eq.~\eqref{eq:matrix_expand} to Eq.~\eqref{eq:quadratic_cost}, this can be represented in the same quadratic form as Eq.~\eqref{QUBOcost}.
Therefore, by applying the conversion rule shown in Eq.~\eqref{QUBOconvert}, $\hat{H}_{\mathrm{P}}$ is written as
\begin{align}
\hat H_{\mathrm P}
&=
\sum_{1\le i<j\le M-1}
\frac{Q_{ij} + Q_{ji}}{2} 
\left(
\hat a_i^\dagger
\hat a_j
+
\hat a_j^\dagger
\hat a_i
\right)
\nonumber\\
&\quad+\sum_{i=1}^{M-1}
Q_{ii}
\hat a_i^\dagger
\hat a_i
\nonumber\\
&\quad+
\frac12
\sum_{i=1}^{M-1}
r_i
\left(
\hat a_i
+
\hat a_i^\dagger
\right)
+
c
\label{eq:problem_hamiltonian_matrix}
\end{align}
Continuous-variable QA is then performed to minimize this Hamiltonian.

\begin{figure}[tbp]
    \centering
    \includegraphics[width=0.75\linewidth]{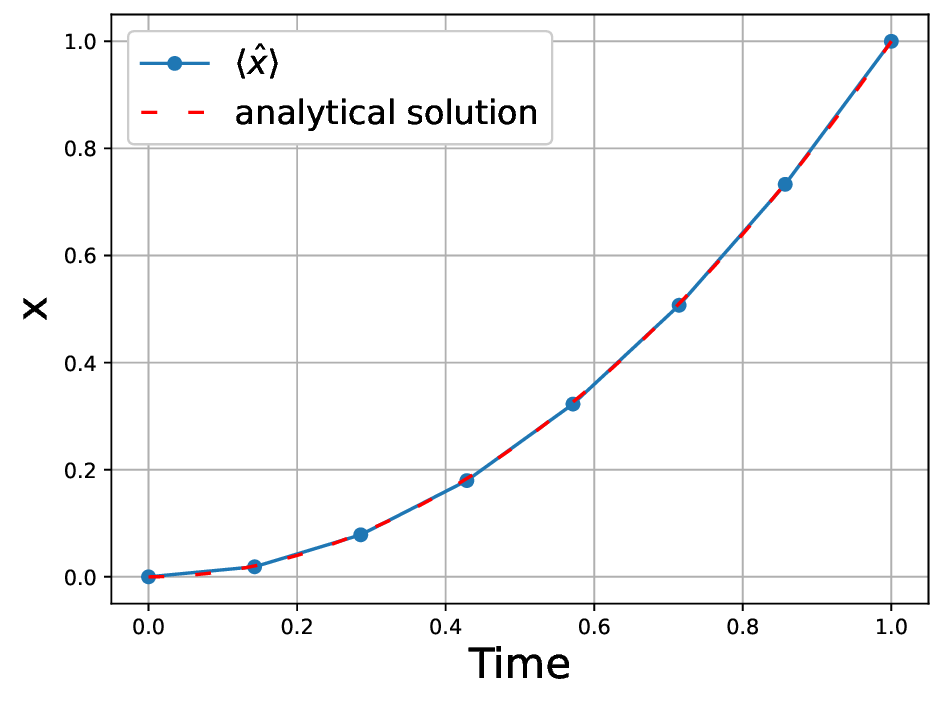}
    \caption{Solution of Eq.~\eqref{eq:example_1}. The horizontal axis represents $\tau$, and the vertical axis represents $x(\tau)$. The numerical results obtained by continuous-variable QA are shown in blue, and the analytical solution is shown by the red dotted line. See Section~\ref{sec:setting} for the parameters used in the numerical calculation.}
    \label{fig:nikaiplot4}
\end{figure}

Figure~\ref{fig:nikaiplot4} compares the solution obtained by the proposed method with the analytical solution $x(\tau)=\tau^2$.
The values of $x$ obtained by QA are close to the analytical solution.
Although $T=10000$ is used here, under the present conditions this annealing time was sufficient to obtain a solution close to the analytical one.
The number of divisions is set to $M=7$, and this moderate number of divisions was sufficient to reproduce the analytical solution in this example.

\begin{figure}[tbp]
    \centering
    \includegraphics[width=0.75\linewidth]{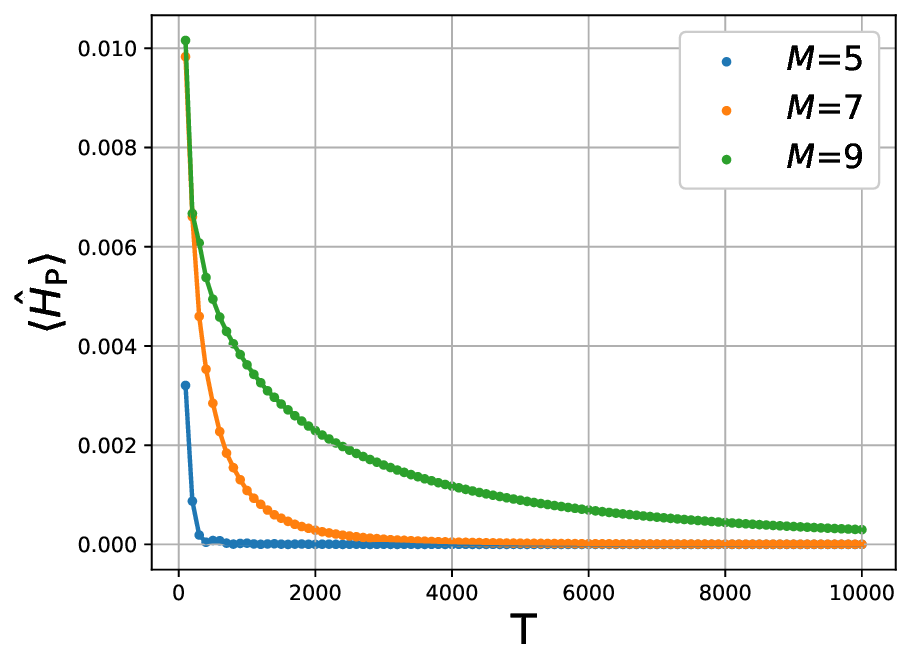}
    \caption{Relationship between the expectation value of $\hat{H}_{\mathrm{P}}$ (vertical axis) and the annealing time $T$ (horizontal axis) when solving Eq.~\eqref{eq:example_1}. As the number of divisions increases, the final expectation value of $\hat{H}_{\mathrm{P}}$ remains farther from zero. Increasing the annealing time $T$ makes the expectation value of $\hat{H}_{\mathrm{P}}$ approach zero. See Section~\ref{sec:setting} for the parameters used in the numerical calculation.}
    \label{fig:nikaiplot4_Tsuii}
\end{figure}

Next, to investigate the properties of continuous-variable QA, we performed QA under several conditions with different annealing times $T$ and different numbers of time divisions $M$.
For each condition, we calculated the expectation value of $\hat{H}_{\mathrm{P}}$ at the end of annealing and evaluated whether it had sufficiently converged to zero.
If the discretized values $x(\tau_i)$ satisfy the finite-difference equations accurately, the expectation value of $\hat{H}_{\mathrm{P}}$ is expected to approach zero.
The results are shown in Fig.~\ref{fig:nikaiplot4_Tsuii}.
The figure shows that increasing the annealing time $T$ decreases the expectation value of $\hat{H}_{\mathrm{P}}$ toward zero.

On the other hand, as the number of divisions $M$ increases, the expectation value of $\hat{H}_{\mathrm{P}}$ does not approach zero as closely, suggesting that the annealing becomes more demanding as the number of unknown variables increases.

\begin{figure}[tbp]
    \centering
    \includegraphics[width=0.75\linewidth]{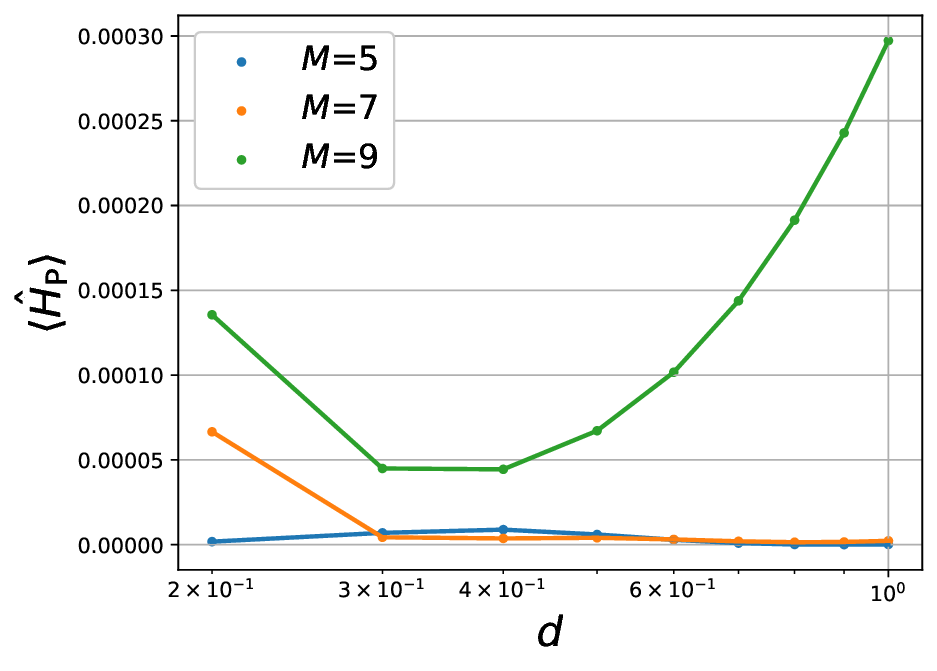}
    \caption{Relationship between the expectation value of $\hat{H}_{\mathrm{P}}$ (vertical axis) and the coefficient $d$ of $\hat{H}_{\mathrm{D}}$ (horizontal axis) when solving Eq.~\eqref{eq:example_1}. Under conditions where annealing succeeds, the expectation value of $\hat{H}_{\mathrm{P}}$ does not change greatly even when $d$ is varied. See Section~\ref{sec:setting} for the parameters used in the numerical calculation.}
    \label{fig:nikai_hd}
\end{figure}

Next, we evaluated the effect of the coefficient $d$ of $\hat{H}_{\mathrm{D}}$ on the results.
Based on the conditions shown in Section~\ref{sec:setting}, Fig.~\ref{fig:nikai_hd} shows the expectation value of $\hat{H}_{\mathrm{P}}$ when the coefficient $d$ of $\hat{H}_{\mathrm{D}}$ is varied.
For $M=5$ and $M=7$, Fig.~\ref{fig:nikai_hd} shows that, in the region where the expectation value of $\hat{H}_{\mathrm{P}}$ is sufficiently small, the expectation value does not vary greatly with changes in the coefficient $d$.
This indicates that, under the conditions where the annealing yields a final state close to the ground state of $\hat{H}_{\mathrm{P}}$, the method is relatively insensitive to the coefficient $d$ of $\hat{H}_{\mathrm{D}}$.
On the other hand, for $M=9$, the expectation value tends to increase as $d$ increases, suggesting that the influence of $d$ becomes more visible for the larger problem size considered here.

\subsection{Application to \texorpdfstring{$\displaystyle \frac{d^2x}{d\tau^2}+5\frac{dx}{d\tau}=-1$}{d2x/dtau2 + 5dx/dtau = -1}}
Using Eqs.~\eqref{sabun_origin} and \eqref{eq:divid_linear_equation} to approximate each derivative term in Eq.~\eqref{eq:example_2} as a linear combination of $x(\tau_i)$, we obtain, for each internal point $i=1,\cdots,M-1$,
\begin{align}
    &\frac{x(\tau_{i+1})-2x(\tau_i)+x(\tau_{i-1})}{(\Delta \tau)^2}
    \nonumber\\
    &\quad+5\frac{x(\tau_{i+1})-x(\tau_{i-1})}{2\Delta \tau} = -1
    \label{after_sabun_2}
\end{align}
Equation~\eqref{after_sabun_2} can also be written in matrix form in the same way as Eq.~\eqref{after_sabun_1}.
The coefficient matrix is
\begin{align}
C_{ij}
=
\begin{cases}
-\dfrac{2}{(\Delta\tau)^2},
&
i=j,
\\[2mm]
\dfrac{1}{(\Delta\tau)^2}
+\dfrac{5}{2\Delta\tau},
&
j=i+1,
\\[2mm]
\dfrac{1}{(\Delta\tau)^2}
-\dfrac{5}{2\Delta\tau},
&
j=i-1,
\\[2mm]
0,
&
\mathrm{otherwise},
\end{cases}
\end{align}
and the vector $\bm{b}$, which combines the initial condition, terminal condition, and constant term, is
\begin{align}
\bm{b}
=
\begin{pmatrix}
-\left(
\dfrac{1}{(\Delta\tau)^2}
-
\dfrac{5}{2\Delta\tau}
\right)
x(\tau_0)
-1
\\
-1
\\
\vdots
\\
-1
\\
-\left(
\dfrac{1}{(\Delta\tau)^2}
+
\dfrac{5}{2\Delta\tau}
\right)
x(\tau_M)
-1
\end{pmatrix}
\label{eq:b_vector_example2}
\end{align}
The subsequent conversion to the quadratic form and the derivation of $\hat{H}_{\mathrm{P}}$ are the same as in Eqs.~\eqref{eq:matrix_cost}--\eqref{eq:problem_hamiltonian_matrix}.
Continuous-variable QA is then performed to minimize this Hamiltonian.

\begin{figure}[tbp]
    \centering
    \includegraphics[width=0.75\linewidth]{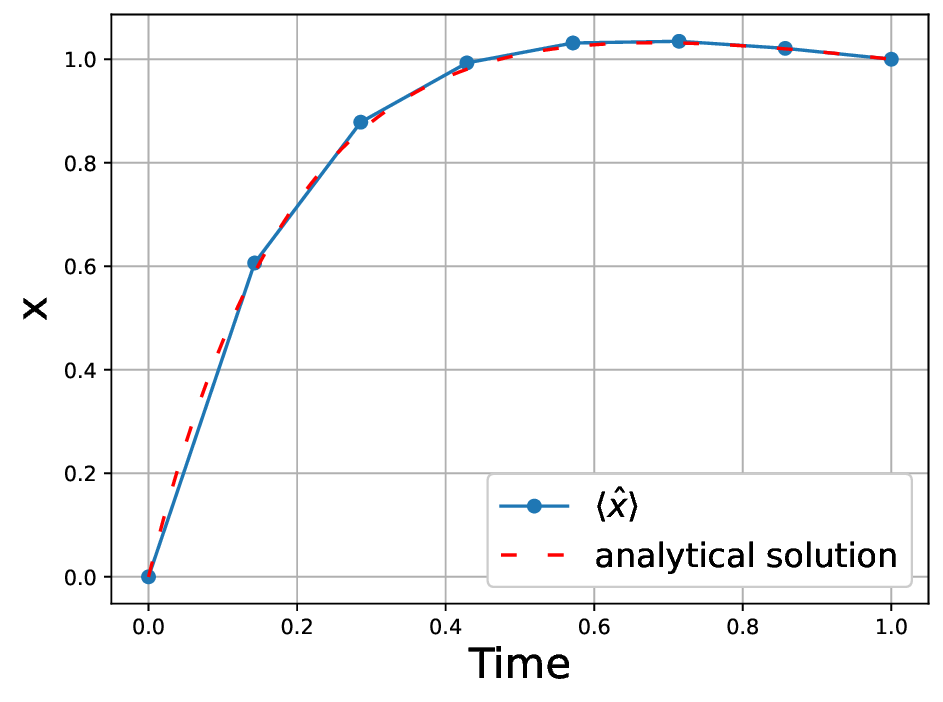}
    \caption{Solution of Eq.~\eqref{eq:example_2}. The horizontal axis represents $\tau$, and the vertical axis represents $x(\tau)$. The numerical results obtained by continuous-variable QA are shown in blue, and the analytical solution is shown by the red dotted line. See Section~\ref{sec:setting} for the parameters used in the numerical calculation.}
    \label{fig:nikaiitikaiplot4}
\end{figure}

Figure~\ref{fig:nikaiitikaiplot4} compares the solution obtained by the proposed method with the analytical solution
$x(\tau)=\frac{6}{5(e^{-5}-1)}(e^{-5\tau}-1)-\frac{\tau}{5}$.
The values of $x$ obtained by QA are close to the analytical solution.
Although $T=10000$ is used here, under the present conditions this annealing time was sufficient to obtain a solution close to the analytical one.
The number of divisions is set to $M=7$, and this moderate number of divisions was sufficient to reproduce the analytical solution in this example.

\begin{figure}[h!]
    \centering
    \includegraphics[width=0.75\linewidth]{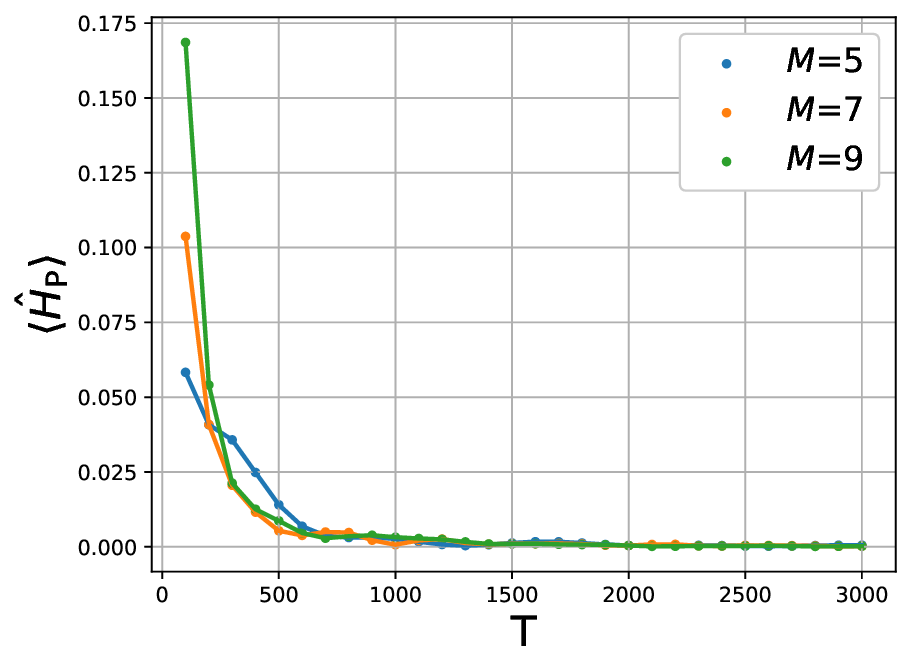}
    \caption{Relationship between the expectation value of $\hat{H}_{\mathrm{P}}$ (vertical axis) and the annealing time $T$ (horizontal axis) when solving Eq.~\eqref{eq:example_2}. Conditions with different numbers of divisions are shown. See Section~\ref{sec:setting} for the parameters used in the numerical calculation.}
    \label{fig:nikaiitikaiplot4_Tsuii}
\end{figure}

Next, to investigate the convergence properties of continuous-variable QA, we performed annealing under several conditions with different annealing times $T$ and different numbers of divisions, and evaluated the expectation value of $\hat{H}_{\mathrm{P}}$ at the end of annealing.
The results are shown in Fig.~\ref{fig:nikaiitikaiplot4_Tsuii}.
As in Fig.~\ref{fig:nikaiplot4_Tsuii}, increasing $T$ makes the expectation value of $\hat{H}_{\mathrm{P}}$ approach zero.
Thus, even when both first- and second-derivative terms are included, the same qualitative convergence behavior is observed.
Conversely, when $T$ is on the order of several hundred, the expectation value of $\hat{H}_{\mathrm{P}}$ does not sufficiently converge to zero.

\begin{figure}[h!]
    \centering
    \includegraphics[width=0.75\linewidth]{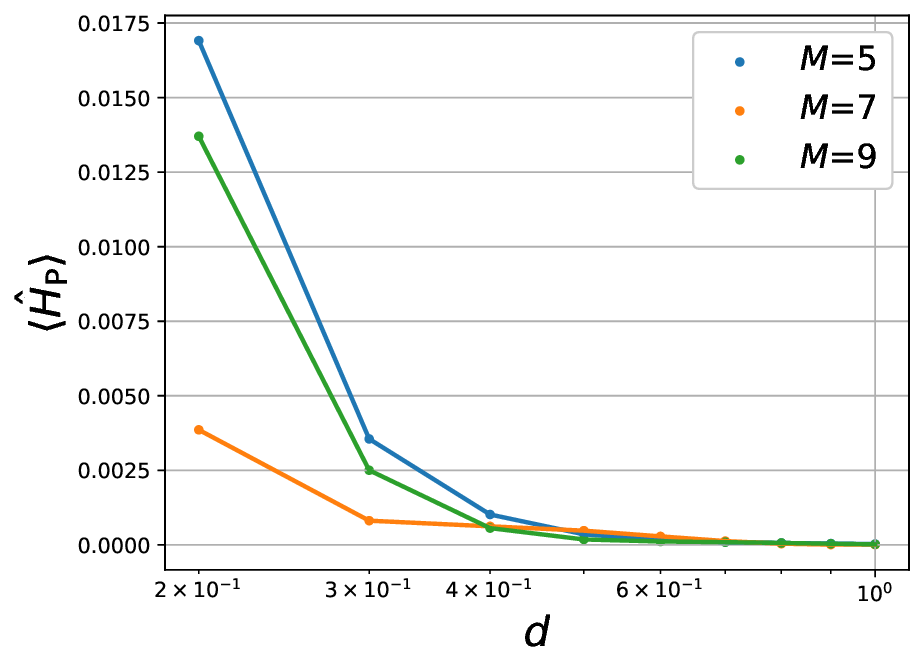}
    \caption{Relationship between the expectation value of $\hat{H}_{\mathrm{P}}$ (vertical axis) and the coefficient $d$ of $\hat{H}_{\mathrm{D}}$ (horizontal axis) when solving Eq.~\eqref{eq:example_2}. Under conditions where annealing succeeds, the expectation value of $\hat{H}_{\mathrm{P}}$ does not change greatly even when $d$ is varied. See Section~\ref{sec:setting} for the parameters used in the numerical calculation.}
    \label{fig:nikaiitikai_hd}
\end{figure}

Figure~\ref{fig:nikaiitikai_hd} shows that, in the region where the expectation value of $\hat{H}_{\mathrm{P}}$ is sufficiently small, the expectation value does not vary greatly with changes in the coefficient $d$.

\section{Conclusion}

In conclusion, we developed a formulation for solving linear differential equations within the recently proposed framework of continuous-variable quantum annealing. Specifically, by discretizing time in a linear differential equation and approximating the derivative terms at each time point using finite differences, we transformed the differential equation into a quadratic optimization problem whose variables can be directly encoded in the continuous degrees of freedom of bosonic cavity modes. Numerical simulations of two second-order linear differential equations demonstrated that, under the conditions examined, the solutions obtained using continuous-variable quantum annealing converge toward the corresponding analytical solutions.

Future work will extend the present formulation to a broader class of linear differential equations, including systems of coupled equations with multiple dependent variables, in order to assess its generality and scalability. It will also be important to investigate methods to improve the accuracy of the solutions obtained at finite annealing times. In particular, nonlinear catalyst Hamiltonians have been shown to enhance ground-state preparation by improving the adiabatic dynamics \cite{koura2024linear,PhysRevE.85.051112}. Examining whether such catalyst terms can similarly improve solution accuracy within the present framework is an important direction for future research.

This project is supported by
JST Moonshot R\&D Grant
Number JPMJMS226C, 
JST CREST Grant Number JPMJCR23I5, and Presto
JST Grant Number JPMJPR245B.

\bibliography{ref}

@article{Kadowaki_1998,
   title={Quantum annealing in the transverse Ising model},
   volume={58},
   ISSN={1095-3787},
   number={5},
   journal={Physical Review E},
   publisher={American Physical Society (APS)},
   author={Kadowaki, Tadashi and Nishimori, Hidetoshi},
   year={1998},
   month=nov, pages={5355–5363} }

@article{finnila1994quantum,
  title={Quantum annealing: A new method for minimizing multidimensional functions},
  author={Finnila, Aleta Berk and Gomez, Maria A and Sebenik, C and Stenson, Catherine and Doll, Jimmie D},
  journal={Chemical physics letters},
  volume={219},
  number={5-6},
  pages={343--348},
  year={1994},
  publisher={Elsevier}
}

@article{farhi2000quantum,
  title={Quantum computation by adiabatic evolution},
  author={Farhi, Edward and Goldstone, Jeffrey and Gutmann, Sam and Sipser, Michael},
  journal={arXiv preprint quant-ph/0001106},
  year={2000}
}

@article{farhi2001quantum,
  title={A quantum adiabatic evolution algorithm applied to random instances of an NP-complete problem},
  author={Farhi, Edward and Goldstone, Jeffrey and Gutmann, Sam and Lapan, Joshua and Lundgren, Andrew and Preda, Daniel},
  journal={Science},
  volume={292},
  number={5516},
  pages={472--475},
  year={2001},
  publisher={American Association for the Advancement of Science}
}

@article{apolloni1989quantum,
  title={Quantum stochastic optimization},
  author={Apolloni, Bruno and Carvalho, C and De Falco, Diego},
  journal={Stochastic Processes and their Applications},
  volume={33},
  number={2},
  pages={233--244},
  year={1989},
  publisher={Elsevier}
}

@article{Lucas2014,
  author  = {Lucas, Andrew},
  title   = {Ising formulations of many {NP} problems},
  journal = {Frontiers in Physics},
  volume  = {2},
  pages   = {5},
  year    = {2014}
}

@article{Hauke2020,
  author  = {Hauke, Philipp and Katzgraber, Helmut G. and Lechner, Wolfgang and Nishimori, Hidetoshi and Oliver, William D.},
  title   = {Perspectives of quantum annealing: methods and implementations},
  journal = {Reports on Progress in Physics},
  volume  = {83},
  number  = {5},
  pages   = {054401},
  year    = {2020},
}

@article{1390001204189531392,
author="Kato, Tosio",
title="On the Adiabatic Theorem of Quantum Mechanics",
journal="Journal of the Physical Society of Japan",
ISSN="00319015",
publisher="一般社団法人 日本物理学会",
year="1950",
volume="5",
number="6",
pages="435-439",
}

@article{Jansen_2007,
   title={Bounds for the adiabatic approximation with applications to quantum computation},
   volume={48},
   ISSN={1089-7658},
   number={10},
   journal={Journal of Mathematical Physics},
   publisher={AIP Publishing},
   author={Jansen, Sabine and Ruskai, Mary-Beth and Seiler, Ruedi},
   year={2007},
   month=oct }

@article{ambainis2006elementaryproofquantumadiabatic,
  author  = {Ambainis, Andris and Regev, Oded},
  title   = {An Elementary Proof of the Quantum Adiabatic Theorem},
  journal = {arXiv preprint quant-ph/0411152},
  year    = {2006}
}

@inproceedings{dema2020support,
  title={Support vector machine for multiclass classification using quantum annealers},
  author={Dema, B and Arai, Junya and Horikawa, Keitarou},
  booktitle={Proc. DEIM Forum},
  year={2020}
}

@article{date2021adiabatic,
  title={Adiabatic quantum linear regression},
  author={Date, Prasanna and Potok, Thomas},
  journal={Scientific reports},
  volume={11},
  number={1},
  pages={21905},
  year={2021},
  publisher={Nature Publishing Group UK London}
}

@article{chancellor2019domain,
  title={Domain wall encoding of discrete variables for quantum annealing and QAOA},
  author={Chancellor, Nicholas},
  journal={Quantum Science and Technology},
  volume={4},
  number={4},
  pages={045004},
  year={2019},
  publisher={IOP Publishing}
}

@article{karimi2019practical,
  title={Practical integer-to-binary mapping for quantum annealers},
  author={Karimi, Sahar and Ronagh, Pooya},
  journal={Quantum Information Processing},
  volume={18},
  number={4},
  pages={94},
  year={2019},
  publisher={Springer}
}

@article{tamura2021performance,
  title={Performance comparison of typical binary-integer encodings in an Ising machine},
  author={Tamura, Kensuke and Shirai, Tatsuhiko and Katsura, Hosho and Tanaka, Shu and Togawa, Nozomu},
  journal={IEEE Access},
  volume={9},
  pages={81032--81039},
  year={2021},
  publisher={IEEE}
}

@article{krakoff2022controlled,
  title={Controlled precision QUBO-based algorithm to compute eigenvectors of symmetric matrices},
  author={Krakoff, Benjamin and Mniszewski, Susan M and Negre, Christian FA},
  journal={Plos one},
  volume={17},
  number={5},
  pages={e0267954},
  year={2022},
  publisher={Public Library of Science San Francisco, CA USA}
}

@article{PhysRevA.108.042403,
  title = {Effectiveness of quantum annealing for continuous-variable optimization},
  author = {Arai, Shunta and Oshiyama, Hiroki and Nishimori, Hidetoshi},
  journal = {Phys. Rev. A},
  volume = {108},
  issue = {4},
  pages = {042403},
  numpages = {21},
  year = {2023},
  month = {Oct},
  publisher = {American Physical Society}
}

@article{koura2024linear,
      author  = {Asuka Koura and Takashi Imoto and Katsuki Ura and Yuichiro Matsuzaki},
  title   = {Linear regression using quantum annealing with continuous variables},
  journal = {Japanese Journal of Applied Physics},
  volume  = {64},
  number  = {3},
  pages   = {03SP31},
  year    = {2025},
  month   = mar
}

@article{bravo2023variational,
  title={Variational quantum linear solver},
  author={Bravo-Prieto, Carlos and LaRose, Ryan and Cerezo, Marco and Subasi, Yigit and Cincio, Lukasz and Coles, Patrick J},
  journal={Quantum},
  volume={7},
  pages={1188},
  year={2023},
  publisher={Verein zur F{\"o}rderung des Open Access Publizierens in den Quantenwissenschaften}
}

@article{sato2023variational,
  title={Variational quantum algorithm for generalized eigenvalue problems and its application to the finite-element method},
  author={Sato, Yuki and Watanabe, Hiroshi C and Raymond, Rudy and Kondo, Ruho and Wada, Kaito and Endo, Katsuhiro and Sugawara, Michihiko and Yamamoto, Naoki},
  journal={Physical Review A},
  volume={108},
  number={2},
  pages={022429},
  year={2023},
  publisher={APS}
}

@article{PhysRevA.104.022418,
  title = {Variational quantum algorithm for the Poisson equation},
  author = {Liu, Hai-Ling and Wu, Yu-Sen and Wan, Lin-Chun and Pan, Shi-Jie and Qin, Su-Juan and Gao, Fei and Wen, Qiao-Yan},
  journal = {Phys. Rev. A},
  volume = {104},
  issue = {2},
  pages = {022418},
  numpages = {12},
  year = {2021},
  month = {Aug},
  publisher = {American Physical Society},
}

@article{PhysRevA.104.052409,
  title = {Variational quantum algorithm based on the minimum potential energy for solving the Poisson equation},
  author = {Sato, Yuki and Kondo, Ruho and Koide, Satoshi and Takamatsu, Hideki and Imoto, Nobuyuki},
  journal = {Physical Review A},
  volume = {104},
  issue = {5},
  pages = {052409},
  numpages = {15},
  year = {2021},
  month = {Nov},
  publisher = {American Physical Society},
}

@article{liu2022application,
  title={Application of a variational hybrid quantum-classical algorithm to heat conduction equation and analysis of time complexity},
  author={Liu, YY and Chen, Zhen and Shu, Chang and Chew, Siou Chye and Khoo, Boo Cheong and Zhao, Xiang and Cui, YD},
  journal={Physics of Fluids},
  volume={34},
  number={11},
  year={2022},
  publisher={AIP Publishing}
}

@article{demirdjian2022variational,
  title={Variational quantum solutions to the advection--diffusion equation for applications in fluid dynamics},
  author={Demirdjian, Reuben and Gunlycke, Daniel and Reynolds, Carolyn A and Doyle, James D and Tafur, Sergio},
  journal={Quantum Information Processing},
  volume={21},
  number={9},
  pages={322},
  year={2022},
  publisher={Springer}
}

@inproceedings{Kudo_2026,
  title     = {Annealing-Based Approach to Solving Partial Differential Equations},
  author    = {Kudo, Kazue},
  booktitle = {Proc. 2026 Int. Conf. Quantum Communications, Networking, and Computing (QCNC)},
  publisher = {IEEE},
  year      = {2026},
  month     = apr,
  pages     = {912--917},
}

@article{PhysRevA.99.052355,
  title = {Box algorithm for the solution of differential equations on a quantum annealer},
  author = {Srivastava, Siddhartha and Sundararaghavan, Veera},
  journal = {Phys. Rev. A},
  volume = {99},
  issue = {5},
  pages = {052355},
  numpages = {10},
  year = {2019},
  month = {May},
  publisher = {American Physical Society},
}

@article{PhysRevA.104.032426,
  title = {Hybrid classical-quantum approach to solve the heat equation using quantum annealers},
  author = {Pollachini, Giovani G. and Salazar, Juan P. L. C. and G\'oes, Caio B. D. and Maciel, Thiago O. and Duzzioni, Eduardo I.},
  journal = {Phys. Rev. A},
  volume = {104},
  issue = {3},
  pages = {032426},
  numpages = {7},
  year = {2021},
  month = {Sep},
  publisher = {American Physical Society},
}

@article{Criado_2023,
year = {2022},
month = {dec},
publisher = {IOP Publishing},
volume = {8},
number = {1},
pages = {015021},
author = {Criado, Juan Carlos and Spannowsky, Michael},
title = {Qade: solving differential equations on quantum annealers},
journal = {Quantum Science and Technology}
}

@article{Qutip,
   title={QuTiP 2: A Python framework for the dynamics of open quantum systems},
   volume={184},
   ISSN={0010-4655},
   number={4},
   journal={Computer Physics Communications},
   publisher={Elsevier BV},
   author={Johansson, J.R. and Nation, P.D. and Nori, Franco},
   year={2013},
   month=Apr, pages={1234–1240} }

@article{qutip5,
  title = {QuTiP 5: The Quantum Toolbox in {Python}},
  author = {
    Lambert, Neill and Gigu{`e}re, Eric and Menczel, Paul and Li, Boxi and
    Hopf, Patrick and Su{'a}rez, Gerardo and Gali, Marc and Lishman, Jake and
    Gadhvi, Rushiraj and Agarwal, Rochisha and Galicia, Asier and Shammah, Nathan and
    Nation, Paul and Johansson, J. R. and Ahmed, Shahnawaz and Cross, Simon and
    Pitchford, Alexander and Nori, Franco
  },
  journal = {Physics Reports},
  volume = {1153},
  pages = {1-62},
  year = {2026},
  issn = {0370-1573},
  
}

@article{PhysRevE.85.051112,
  title = {Quantum annealing with antiferromagnetic fluctuations},
  author = {Seki, Yuya and Nishimori, Hidetoshi},
  journal = {Phys. Rev. E},
  volume = {85},
  issue = {5},
  pages = {051112},
  numpages = {8},
  year = {2012},
  month = {May},
  publisher = {American Physical Society},
}

\end{document}